\documentclass{article}
\usepackage[preprint]{neurips_2024}
\usepackage[utf8]{inputenc}
\usepackage[T1]{fontenc}
\usepackage{amsmath,amssymb}
\usepackage{booktabs}
\usepackage{graphicx}
\usepackage{hyperref}
\hypersetup{hidelinks}
\usepackage{natbib}
\usepackage{xcolor}
\usepackage{multirow}
\usepackage{subcaption}

\title{Emergent Compositional Communication\\for Latent World Properties}

\author{
  Tomek Kaszy\'{n}ski \\
  Independent Researcher, Amsterdam, Netherlands \\
  \texttt{t.kaszynski@proton.me} \\
  \url{https://github.com/TomekKaszynski/emergent-physics-comm}
}

\date{March 2026}

\begin{document}
\maketitle

% ================================================================
\begin{abstract}
What physical knowledge do video foundation models encode, and can it be extracted into discrete, compositional form? We show that multi-agent communication pressure, combined with a discrete Gumbel-Softmax bottleneck and iterated learning, induces compositional representations of world properties that are invisible in any single observation---elasticity, friction, mass ratio---from frozen pretrained features alone.

\textbf{The backbone determines what is communicable.} In a controlled $2{\times}2$ factorial comparison on physics simulations, DINOv2 \citep{oquab2023dinov2} dominates on spatially-visible physics (ramp: 98.3\% vs.\ 95.1\%), while V-JEPA~2 \citep{assran2025vjepa2} dominates on dynamics-only collision physics where properties are recoverable only from temporal velocity differences (87.4\% vs.\ 77.7\%, $d{=}2.74$). Scale-matched and frame-matched controls definitively attribute this gap to video-native pretraining: DINOv2 ViT-L at matched parameters performs \textit{worse} ($d{=}3.37$), and DINOv2 with matched frame count degrades further ($d{=}6.53$). This extends the findings of \citet{garrido2025intuitive}, who showed V-JEPA representations encode intuitive physics: our results demonstrate that this physical knowledge can be compressed into discrete, compositional codes under communication pressure.

\textbf{Multi-agent structure drives compositionality.} With 4 agents, 100\% of seeds converge to near-perfect positional disentanglement (PosDis $= 0.999$, holdout 98.3\%; $n{=}80$), while 2 agents produce compositional protocols stochastically (54\%). Targeted controls confirm the driver is multi-agent structure---not bandwidth or temporal coverage. Causal intervention via position-zeroing shows surgical property disruption ($\sim$15\% drop on targeted property, $<$3\% on others), and the frozen protocol transfers to cross-property reasoning (93.8\%), outcome prediction (88.7\% at 25$\times$ compression), and action-conditioned planning with counterfactual velocity reasoning (91.5\%, $r{=}0.780$). Validation on Physics~101 real camera footage \citep{wu2016physics101} confirms mass-comparison accuracy of 85.6\% on unseen objects, with temporal dynamics contributing +11.2\% beyond static appearance, agent-scaling compositionality replicating at 90\% for 4 agents, and the causal intervention extending to real video (zeroing the mass-relevant agent reduces accuracy by 7.8pp while the other causes only 2.1pp disruption; $p{=}0.022$, $d{=}1.87$). The discrete communication channel functions as an empirical instantiation of the latent variable $z$ in \citeauthor{lecun2022path}'s (\citeyear{lecun2022path}) cognitive architecture---our results provide evidence that discrete latent structure within JEPA-style world models supports compositional, causally addressable physical reasoning.
\end{abstract}

% ================================================================
\section{Introduction}

Physical understanding requires inferring properties that are never directly visible. Elasticity, friction, and damping cannot be observed in a photograph---they reveal themselves only through temporal dynamics: how objects bounce, slide, and decelerate. Humans infer these properties effortlessly and communicate about them using compositional language (``the bouncy, slippery ball''). Can artificial agents develop similar capabilities from perceptual experience, driven by the need to communicate? Understanding how agents develop structured representations of latent physical properties has implications for robotics (communicating material properties between manipulators), interpretable world models (discrete, inspectable representations of physical concepts), and multi-agent coordination in partially observable environments.

We study this question at the intersection of emergent communication \citep{lazaridou2017multi, havrylov2017emergence, lazaridou2020emergent} and intuitive physics \citep{battaglia2013simulation, piloto2022intuitive}. Prior work in emergent communication has focused on referential games with directly visible attributes---color, shape, position \citep{choi2018compositional, chaabouni2020compositionality, ren2020compositional}. Work on physical understanding from video has focused on prediction tasks with continuous representations \citep{wu2017learning, ye2018interpretable}. We are not aware of prior work that studies emergent discrete communication about latent physical properties inferred from temporal dynamics, with explicit compositional generalization holdouts.

Our setup is simple. Two agents each observe video of a ball interacting with a ramp. Each ball has invisible physical properties---elasticity and friction---drawn from a $5{\times}5$ grid. The agents must cooperate on a pairwise comparison task: determine which ball has higher elasticity and which has higher friction. They communicate through a discrete bottleneck: a short message of categorical symbols transmitted via Gumbel-Softmax \citep{jang2017categorical}. Each agent observes only its own scene---the bottleneck forces compression of physical information into discrete symbols. Senders never observe property values and never receive supervision on message structure; learning is driven entirely by task loss at the receiver.

The key question is whether the structure of the message reflects the structure of the world. Do agents develop \textit{compositional} protocols where different message positions encode different properties?

We find that given three conditions---(1) a perceptual backbone capable of extracting relevant properties (DINOv2; \citealt{oquab2023dinov2}), (2) a factored message structure providing positional slots, and (3) iterated learning with population-based receiver pressure \citep{kirby2014iterated, rita2022emergent}---over half of 2-agent training runs spontaneously develop positionally disentangled protocols ($n{=}80$, 54\% compositional, with significantly better generalization, $p{=}0.011$). Scaling to 4 agents yields compositionality in all 80 seeds tested (PosDis $= 0.999$, holdout 98.3\%). Causal intervention confirms this is genuine compositionality: zeroing a single message position selectively disrupts the property it encodes while leaving others intact (\S\ref{sec:intervention}).

Our five main contributions are:

\begin{enumerate}
    \item \textbf{Compositional communication about invisible physics.} Agents develop positionally disentangled protocols from raw video (54\% of 80 seeds with 2 agents; 100\% with 3 and 4 agents across all seeds tested). Targeted controls show that this transition is driven by multi-agent structure, not temporal coverage or bandwidth: randomizing frame assignments preserves the effect, while matched-bandwidth single senders do not (\S\ref{sec:scaling}). Compositionality provides a significant generalization advantage ($p{=}0.011$, Cohen's $d{=}0.59$), and a frozen compositional sender enables cross-property reasoning on a novel task at 93.8\% accuracy. An alternative compositionality pressure (LazImpa) achieves 0\% compositionality, confirming iterated learning is necessary in our experiments (\S\ref{sec:compositionality}--\ref{sec:intervention}).
    \item \textbf{Perception determines what is communicable.} In a controlled $2{\times}2$ factorial comparison (two backbones $\times$ two datasets, matched 4-agent configurations), V-JEPA~2 significantly outperforms DINOv2 on collision dynamics (87.4\% vs.\ 77.7\%, $p{<}0.0001$), while DINOv2 significantly outperforms V-JEPA~2 on ramp physics (98.3\% vs.\ 95.1\%, $p{=}0.001$). A scale-matched control (DINOv2 ViT-L, 304M parameters matching V-JEPA~2) confirms the collision gap is driven by video pretraining, not model capacity: DINOv2 ViT-L achieves only 74.6\% ($d{=}3.37$ vs.\ V-JEPA~2). At 2 agents, no backbone difference is detectable on either dataset, revealing that the discrete bottleneck equalizes performance until sufficient communication bandwidth exposes backbone-level encoding differences (\S\ref{sec:backbone}).
    \item \textbf{Consistent mechanism across domains with information-driven specialization.} The identical communication architecture produces compositional protocols across four domains. Agent specialization patterns reorganize to match each domain's information structure, ranging from near-perfect disentanglement (specialization ratio 0.95 on spring-mass) to distributed encoding (0.20 on 6-property vision). Agents allocate bandwidth proportional to property extractability ($r{=}0.964$, $n{=}6$), consistent with rate-distortion principles \citep{cover2006elements} (\S\ref{sec:domains}--\ref{sec:bandwidth}).
    \item \textbf{Boundary condition: continuous vs.\ categorical.} The mechanism succeeds on continuous property variation but fails on categorical visual recognition, where agents memorize class codes. This identifies a boundary condition for when compositional structure emerges under communication pressure (\S\ref{sec:boundary}).
    \item \textbf{Real-video validation.} We validate the mechanism on Physics~101 \citep{wu2016physics101}, a real-camera dataset with laboratory-measured mass labels. Agents achieve 85.6\% mass-comparison accuracy on unseen objects, with temporal dynamics contributing +11.2\% beyond static appearance. The agent-scaling effect replicates on real video: 4-agent compositionality reaches 90\% with two physical properties (mass and derived restitution), confirming that multi-agent pressure induces compositional structure beyond synthetic environments (\S\ref{sec:realvideo}).
\end{enumerate}

% ================================================================
\section{Related Work}

\paragraph{Emergent Communication.}
The study of language emergence in multi-agent systems has grown rapidly since \citet{lazaridou2017multi} and \citet{havrylov2017emergence}. \citet{lazaridou2020emergent} provide a comprehensive survey of this field. A central question is whether emergent protocols exhibit compositionality---systematic structure where the meaning of a message is determined by the meanings of its parts \citep{brighton2006understanding, andreas2019measuring}. \citet{kottur2017natural} showed that compositionality requires restricted channel capacity. \citet{chaabouni2020compositionality} introduced positional disentanglement (PosDis) as a metric and found that smaller vocabularies promote compositional structure. \citet{ren2020compositional} demonstrated compositionality in visual referential games. However, \citet{kharitonov2020emergent} showed that compositionality and generalization are not always tightly linked, highlighting the need for careful evaluation. Grounded compositional language has been studied in multi-agent populations by \citet{mordatch2018emergence}, though in settings with directly observable attributes. In the multi-agent reinforcement learning literature, \citet{das2019tarmac} introduced targeted multi-agent communication where agents learn what to communicate and whom to address---our work shares the premise that communication structure matters, but studies how compositional protocols emerge rather than how to route messages.

Several mechanisms encourage compositionality. Iterated learning---periodically resetting the receiver---creates pressure for learnable protocols \citep{kirby2014iterated, li2019ease, ren2020compositional, rita2022emergent}. Population-based training forces universal decodability \citep{tieleman2019shaping}. Factored message structures provide positional slots for specialization \citep{chaabouni2020compositionality}. \citet{rita2020lazimpa} introduced an alternative: ``lazy'' speakers (penalizing message entropy) with ``impatient'' listeners (predicting from partial messages). We compare against this approach and find it insufficient for latent property communication. We combine iterated learning, population pressure, and factored messages, and apply them to a novel domain: latent physical properties that must be inferred from temporal dynamics, requiring genuine physical understanding rather than pattern matching on static features.

\paragraph{Physical Understanding from Video.}
Learning physical properties from visual observations has been studied through prediction-centric approaches. \citet{wu2017learning} inferred physical properties via analysis-by-synthesis. \citet{bear2021physion} introduced benchmarks for physical prediction. \citet{piloto2022intuitive} showed that deep learning systems can develop intuitive physics knowledge. These approaches use continuous latent representations. Our work differs by requiring agents to \textit{communicate} about physics through discrete symbols, creating pressure for categorical abstraction of continuous physical quantities. Physics~101 \citep{wu2016physics101} provides real-camera video of 101 objects with measured mass and volume across controlled physics scenarios; we use it for real-video validation of our emergent communication mechanism (\S\ref{sec:realvideo}).

\paragraph{Vision Foundation Models.}
DINOv2 \citep{oquab2023dinov2} provides general-purpose visual features through self-supervised learning on images. V-JEPA~2 \citep{assran2025vjepa2} extends self-supervised learning to video, using a joint-embedding predictive architecture trained on over one million hours of internet video to capture temporal dynamics and motion understanding. Recent evaluation shows that V-JEPA representations support intuitive physics understanding (object permanence, shape constancy) but struggle with precise collision dynamics \citep{garrido2025intuitive}---a finding our collision experiments directly address. We use frozen features from both backbones: DINOv2 provides per-frame spatial features while V-JEPA~2 provides spatiotemporal tokens encoding motion. The communication bottleneck decomposes these holistic features into compositional descriptions---a transformation neither backbone performs alone. By comparing both backbones on tasks with different temporal demands, we probe how the perceptual prior constrains emergent communication.

% ================================================================
\section{Method}

\subsection{Environment and Task}

\paragraph{Ramp Physics (Primary).} We generate physics simulations using Kubric \citep{greff2022kubric} with PyBullet. A sphere slides down a 70\textdegree{} ramp (revealing friction through slide speed) and bounces on flat ground (revealing elasticity through bounce height). Scenes are rendered at $128{\times}128$ RGB, 24 frames at 12 fps. Properties are varied on a $5{\times}5$ grid: elasticity $\in \{0.1, 0.3, 0.5, 0.7, 0.9\}$ and friction $\in \{0.1, 0.3, 0.5, 0.7, 0.9\}$, yielding 25 combinations with 12 scenes each (300 total). Visual nuisance variables---ball color, lighting, spawn position---are randomized independently.

\paragraph{Collision Dynamics (Backbone Comparison).} To test whether the perceptual backbone constrains what agents can communicate about, we design a second dataset where temporal reasoning is strictly required. Two \textit{visually identical} spheres (same size, color, texture) collide on a flat surface. Sphere~A approaches at randomized velocity; Sphere~B is stationary. Hidden properties are mass ratio ($m_B/m_A \in \{1, 2, 3, 4, 5\}$) and collision elasticity (restitution $\in \{0.1, 0.3, 0.5, 0.7, 0.9\}$). Since both spheres are visually indistinguishable, mass ratio cannot be inferred from any single frame---agents must track post-collision velocity differences over time. Scenes are rendered at $256{\times}256$ RGB, 48 frames at 24 fps (2 seconds). The $5{\times}5$ grid yields 600 scenes (24 per cell). DINOv2 extracts CLS tokens from 24 evenly-spaced frames independently; V-JEPA~2 processes all 48 frames jointly and its output is subsampled to 24 temporal positions (see \S\ref{sec:training} for details).

\paragraph{Task.} Two sender agents $S_A$ and $S_B$ each observe an input $x_i$ and produce a discrete message $m_i$. A receiver $R$ observes both messages and must determine, for each property: which input has the higher value? The loss is the sum of binary cross-entropy across all properties. Each sender observes only its own input. Senders never see property values---they receive gradients only through the receiver's task loss.

\paragraph{Compositionality Holdout.} To test compositional generalization, we hold out specific property \textit{combinations} while ensuring every individual value appears in training. A Latin square removes 5 cells from the $5{\times}5$ grid (one per row and column), yielding 240 training and 60 test scenes. An agent encoding each property independently should generalize to novel combinations; an agent memorizing specific combinations should not.

\subsection{Architecture}

\paragraph{Perception.} For ramp physics, we use frozen DINOv2 ViT-S/14 \citep{oquab2023dinov2} CLS tokens from 8 evenly-spaced frames, yielding $(8, 384)$ temporal features. For collision dynamics, we compare two backbones under identical downstream architecture: (1) DINOv2 ViT-S/14 CLS tokens extracted independently from 24 evenly-spaced frames, yielding $(24, 384)$; (2) V-JEPA~2 ViT-L/16 \citep{assran2025vjepa2}, which ingests all 48 frames jointly through its spatiotemporal transformer, producing 6{,}144 spatiotemporal tokens that we reshape to $(24, 256, 1024)$ (24 temporal positions $\times$ 256 spatial tokens) and spatially average to $(24, 1024)$. \textbf{Note:} V-JEPA~2 sees all 48 raw frames; the ``24'' in its output shape reflects temporal positions after the model's internal temporal stride, not subsampling at the input level. Both representations are fed through a temporal encoder (two 1D convolutions with ReLU and adaptive average pooling) to produce a 128-dimensional scene representation. This provides a controlled comparison: identical temporal aggregation and downstream architecture, different perceptual features.

\paragraph{Sender.} The sender maps the scene representation to a discrete message through $K$ independent Gumbel-Softmax heads \citep{jang2017categorical}, each with vocabulary size $V$:
\begin{equation}
    m = [\text{GS}(W_1 h),\; \text{GS}(W_2 h),\; \ldots,\; \text{GS}(W_K h)]
\end{equation}
where $h$ is the scene representation and GS denotes Gumbel-Softmax with temperature annealing ($\tau: 2.0 \to 0.5$). The main configuration uses $K{=}2, V{=}5$ (25 possible messages). The factored structure provides the \textit{capacity} for positional specialization; whether agents \textit{use} it compositionally depends on training dynamics.

\paragraph{Receiver.} The receiver concatenates one-hot messages from both senders and processes them through a shared MLP trunk ($2 \cdot K \cdot V \to 128 \to 64$, ReLU) with separate sigmoid output heads per property comparison.

\paragraph{Terminology.} To avoid ambiguity: an \textit{agent} is a sender module with its own encoder and message head, observing a subset of the input frames. A \textit{message position} is one of the $K$ independent Gumbel-Softmax heads within a single agent's message. \textit{Bandwidth} refers to the total number of discrete symbols transmitted per input: with $N$ agents, each sending $K$ positions of vocabulary $V$, the total is $N \cdot K$ symbols (from a space of $V^{N \cdot K}$ possible joint messages). In the ``matched bandwidth'' control (\S\ref{sec:scaling}), a single sender ($N{=}1$) transmits $K{=}4$ positions of $V{=}5$, matching the $N{=}4, K{=}2, V{=}5$ multi-agent condition in total symbols ($4 \times 5 = 20$ one-hot values) while removing the structural constraint of independent partial observers.

\paragraph{Holistic Control.} A single Gumbel-Softmax head with vocabulary $V^K$ (identical channel capacity, no positional structure).

\subsection{Training}
\label{sec:training}

\paragraph{Oracle Pretraining.} An oracle model with the same encoder directly compares two inputs without communication. We pretrain for 100 epochs, then copy encoder weights to initialize the sender.

\paragraph{Population-Based Iterated Learning.} Following \citet{rita2022emergent}, we train the sender against a population of 3 receivers simultaneously. The sender loss averages across all receivers, forcing universally decodable messages. Every 40 epochs, all receivers are reset simultaneously, creating maximum learnability pressure---the sender's language must be learnable from scratch by any new receiver. This simulates cultural transmission \citep{kirby2014iterated}: protocols that are easier to learn (i.e., more compositional) are favored. Over 400 epochs, this produces 9 receiver ``generations.'' Asymmetric learning rates (sender: $10^{-3}$, receivers: $3 \times 10^{-3}$) ensure receivers adapt faster, so the sender's protocol is the stable element while receivers must conform to it.

\paragraph{Regularization.} Entropy regularization ($-0.03 \cdot H$) activates when per-head symbol entropy drops below $0.1 \cdot \log V$, preventing vocabulary collapse. Gradient clipping at 1.0 handles Gumbel-Softmax instabilities.

\subsection{Metrics}

\paragraph{Positional Disentanglement (PosDis).} Introduced by \citet{chaabouni2020compositionality}. For each message position $k$, we compute mutual information $\text{MI}(m_k; a_j)$ with each attribute $a_j$. A high PosDis means each position predominantly encodes one property and minimally encodes others:
\begin{equation}
    \text{PosDis} = \frac{1}{K} \sum_{k=1}^{K} \frac{\max_j \text{MI}(m_k; a_j) - \text{second-max}_j \text{MI}(m_k; a_j)}{\max_j \text{MI}(m_k; a_j) + \epsilon}
\end{equation}

\paragraph{Topographic Similarity (TopSim).} Spearman correlation between meaning distances (Manhattan distance in property space) and message distances (Hamming distance) \citep{brighton2006understanding}.

\paragraph{Bag-of-Symbols Disentanglement (BosDis).} Like PosDis but position-independent: measures whether specific symbols (regardless of which position they appear in) uniquely encode specific attributes. High BosDis confirms specialization is robust to position assumptions.

% ================================================================
\section{Experiments}

All experiments use the architecture and training described in \S\ref{sec:training}. We report means and standard deviations across 20 seeds per condition (80 seeds for the main 2-agent ramp characterization in \S\ref{sec:compositionality}). Full hyperparameters appear in Appendix~\ref{app:hparams}.

\subsection{Compositional Communication About Invisible Physics}
\label{sec:compositionality}

\begin{figure}[h]
\centering
\includegraphics[width=0.7\textwidth]{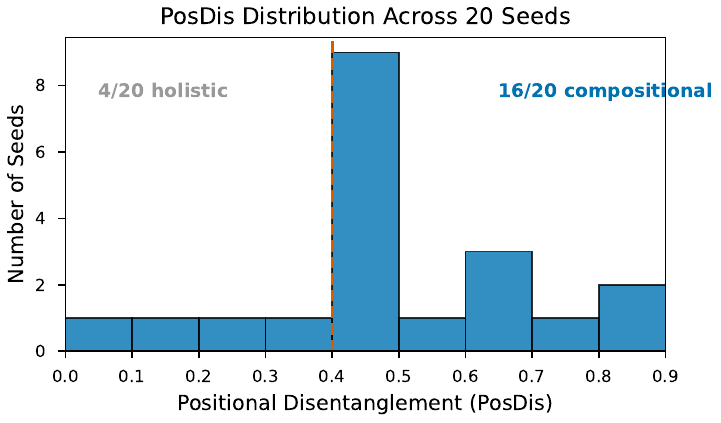}
\caption{PosDis distribution across 20 seeds (representative subset; full 80-seed characterization in Table~\ref{tab:main}). Compositional seeds (PosDis $> 0.4$) generalize significantly better than holistic seeds. The distribution is bimodal with mass above 0.4 and below 0.3; the same pattern holds across all 80 seeds (43/80 compositional).}
\label{fig:phase_transition}
\end{figure}

With $2{\times}5$ messages (two positions, five symbols each) and DINOv2 features, population-based iterated learning over 400 epochs produces compositional protocols in 54\% of seeds (43/80 with PosDis $> 0.4$). The PosDis distribution is bimodal (mean 0.444, median 0.434): seeds cluster either above 0.4 or below 0.3, with a clear gap between compositional and holistic regimes (Figure~\ref{fig:phase_transition}). We use 0.4 as the compositionality threshold following this bimodal gap; results are qualitatively unchanged at thresholds of 0.3 or 0.5. Compositional seeds generalize significantly better than holistic ones on unseen property combinations (79.3\% vs.\ 75.9\%, $p{=}0.011$, Cohen's $d{=}0.59$), and PosDis correlates with holdout accuracy ($r{=}0.43$, $p{=}0.0001$).

\begin{table}[h]
\centering
\caption{Two-property communication on ramp physics (80 seeds, 400 epochs). Holdout accuracy on unseen elasticity-friction combinations. Compositional seeds (PosDis $> 0.4$) generalize significantly better ($p{=}0.011$, Cohen's $d{=}0.59$).}
\label{tab:main}
\begin{tabular}{lcccc}
\toprule
Group & $N$ & Holdout (both) & PosDis & Comp.\ seeds \\
\midrule
$2{\times}5$ compositional (PosDis $> 0.4$) & 43 & 79.3\% $\pm$ 5.2\% & 0.617 $\pm$ 0.15 & 43/80 (54\%) \\
$2{\times}5$ holistic (PosDis $\leq 0.4$) & 37 & 75.9\% $\pm$ 6.1\% & 0.244 $\pm$ 0.09 & --- \\
$1{\times}25$ control & 20 & 71.2\% $\pm$ 3.5\% & 0.000 & 0/20 (0\%) \\
\bottomrule
\end{tabular}
\end{table}

\begin{figure}[h]
\centering
\includegraphics[width=\textwidth]{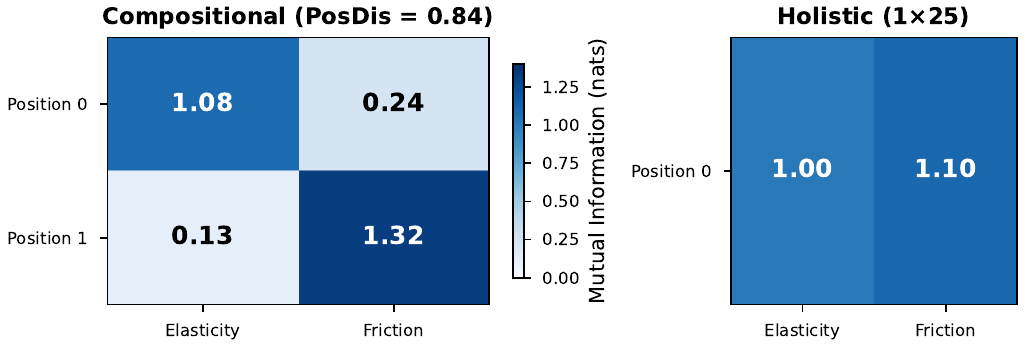}
\caption{Mutual information between message positions and physical properties. Compositional agents (left) show clean diagonal specialization; holistic agents (right) encode both properties in a single undifferentiated symbol.}
\label{fig:mi_heatmap}
\end{figure}

The MI matrix for compositional seeds shows clean position-to-property specialization: Position 0 encodes elasticity (MI $= 1.08$) while Position 1 encodes friction (MI $= 1.32$), with minimal cross-encoding (Figure~\ref{fig:mi_heatmap}).

\paragraph{Ablation Progression.} Table~\ref{tab:ablation} shows how each training ingredient contributes. Without iterated learning, agents overfit heavily (train 93.6\% vs.\ holdout 77.7\%, gap 15.9pp, paired $t{=}10.07$, $p{<}0.0001$) and only 20\% develop compositional protocols. An alternative compositionality pressure---LazImpa \citep{rita2020lazimpa}, combining a lazy speaker (entropy penalty $\lambda{=}0.01$) with an impatient listener (per-position receiver heads)---achieves 0\% compositionality and 70.3\% holdout accuracy. We note that LazImpa was originally designed to promote \textit{efficiency} (Zipf's Law of Abbreviation) rather than compositionality per se; its failure here indicates that efficiency pressure alone, without iterated learning's cultural transmission bottleneck, is insufficient for latent property communication. Furthermore, scaling from 2 to 4 agents eliminates stochastic failure entirely: 100\% of 80 seeds achieve near-perfect compositionality (PosDis $= 0.999$), with holdout accuracy rising from 79.3\% to 98.3\% (\S\ref{sec:scaling}).

\begin{table}[h]
\centering
\caption{Ablation of training components (all using DINOv2, $2{\times}5$ vocabulary). Train-holdout gap reveals that iterated learning regularizes: no-IL overfits heavily while IL sacrifices train accuracy for compositional structure. LazImpa \citep{rita2020lazimpa}, designed for efficiency rather than compositionality, does not produce disentangled protocols on this task. Scaling to 4 agents makes compositionality reliable across all seeds. End-to-end fine-tuning of the encoder degrades communication despite improving feature quality ($R^2 = 0.988$ vs.\ $0.953$).}
\label{tab:ablation}
\begin{tabular}{lccccc}
\toprule
Configuration & Epochs & Train & Holdout & PosDis & Comp.\ \% \\
\midrule
LazImpa (no IL, lazy+impatient) & 400 & --- & 70.3\% & 0.165 & 0\% \\
No iterated learning & 400 & 93.6\% & 77.7\% & 0.311 & 20\% \\
IL + population (simultaneous) & 200 & --- & 74.5\% & 0.295 & 40\% \\
\textbf{IL + pop.\ + extended (2 agents)} & \textbf{400} & --- & \textbf{79.3\%} & \textbf{0.444} & \textbf{54\%} \\
\textbf{IL + pop.\ + extended (4 agents)} & \textbf{400} & --- & \textbf{98.3\%} & \textbf{0.999} & \textbf{100\%} \\
\midrule
\multicolumn{6}{l}{\textit{Encoder ablation (2 agents, IL + pop., 400 epochs):}} \\
E2E fine-tuned DINOv2 & 400 & --- & 67.8\% & 0.415 & 40\% \\
\bottomrule
\end{tabular}
\end{table}

\paragraph{Metric Corroboration.} Complementary compositionality metrics confirm the PosDis findings. For the 80-seed 2-agent characterization, TopSim correlates with PosDis ($r{=}0.669$) and with holdout accuracy ($r{=}0.43$, matching the PosDis--accuracy correlation). Compositional seeds (PosDis $> 0.4$) achieve significantly higher TopSim than holistic seeds. At the 4-agent level, all three metrics converge: on the ramp dataset with DINOv2, TopSim $= 0.792$ and BosDis $= 0.993$ alongside PosDis $= 0.999$. On the collision dataset, V-JEPA~2 achieves TopSim $= 0.738$ and BosDis $= 0.948$ (PosDis $= 0.962$), while DINOv2 ViT-S achieves TopSim $= 0.610$ and BosDis $= 0.855$ (PosDis $= 0.904$). The backbone gap is consistent across all metrics.

\paragraph{Perceptual Bottleneck.} A standard 4-layer CNN encoder achieves 60.6\% on elasticity but only chance on friction---friction manifests as subtle slide-speed variations invisible to shallow convolutions but captured by DINOv2's pretrained features. This demonstrates that the communication bottleneck can only structure what the encoder can extract: adequate perception is a prerequisite, not something communication pressure creates. Conversely, end-to-end fine-tuning of DINOv2 \textit{degrades} communication despite producing objectively better features: E2E achieves higher probe $R^2$ (0.988 vs.\ 0.953) but significantly lower holdout accuracy (67.8\% vs.\ 78.0\%, $p{=}0.0002$, Cohen's $d{=}1.35$, $n{=}20$ per condition) and a lower compositionality rate (40\% vs.\ 60\%). The frozen encoder's information constraint forces the discrete bottleneck to do the structuring work; when the encoder can fine-tune, it bypasses the bottleneck, improving training fit but hurting generalization.

\paragraph{Failure Mode Characterization.} The 37 non-compositional seeds (PosDis $\leq 0.4$) exhibit \textit{holistic} failure: both message positions encode both properties roughly equally, rather than one position collapsing. Analysis of a 20-seed subset confirms vocabulary usage is comparable to compositional seeds (mean entropy 0.966 vs.\ 0.962, $p = 0.60$), ruling out symbol collapse. Holdout accuracy is modestly lower (75.9\% vs.\ 79.3\%, $p = 0.011$)---holistic codes are functional but not factored. Failure shows no correlation with seed index, confirming it is stochastic: some initializations land in a holistic basin from which iterated learning cannot escape within 400 epochs.

\subsection{Causal Evidence: Cross-Property Reasoning}
\label{sec:intervention}

The strongest evidence that the protocol is genuinely compositional---not just correlated with properties---comes from a novel cross-property task. We freeze a compositional sender (PosDis $= 0.753$) trained on same-property comparison (``which ball has higher elasticity?'') and test it on a task it was \textit{never trained on}: ``Is ball A's elasticity greater than ball B's friction?'' This requires the receiver to extract elasticity from message A and friction from message B---different properties from different messages.

\begin{table}[h]
\centering
\caption{Cross-property reasoning (20 seeds). The frozen compositional sender outperforms both freshly-trained conditions on a novel task.}
\label{tab:cross}
\begin{tabular}{lcc}
\toprule
Condition & Holdout & Std \\
\midrule
Frozen compositional sender & \textbf{93.8\%} & 0.7\% \\
Freshly-trained $2{\times}5$ & 92.2\% & 2.4\% \\
Freshly-trained $1{\times}25$ holistic & 87.5\% & 9.5\% \\
\bottomrule
\end{tabular}
\end{table}

\paragraph{Surgical Ablation.} We zero individual message positions and measure accuracy drop. If the protocol is compositional, zeroing the elasticity position of message A should hurt, while zeroing the friction position of message A should not (since friction of A is irrelevant to the task).

\begin{table}[h]
\centering
\caption{Position ablation for the frozen compositional sender on cross-property task. Zeroing relevant positions causes large drops; irrelevant positions cause negligible drops.}
\label{tab:ablation_positions}
\begin{tabular}{lcc}
\toprule
Zeroed Position & Accuracy Drop & Relevance \\
\midrule
A position 0 (elasticity of A) & $-$14.7\% $\pm$ 1.6\% & \textbf{Relevant} \\
A position 1 (friction of A) & $-$0.4\% $\pm$ 1.9\% & Irrelevant \\
B position 0 (elasticity of B) & $-$2.9\% $\pm$ 1.7\% & Irrelevant \\
B position 1 (friction of B) & $-$15.2\% $\pm$ 0.8\% & \textbf{Relevant} \\
\bottomrule
\end{tabular}
\end{table}

The receiver performs surgical extraction: it reads elasticity from position 0 of message A and friction from position 1 of message B, ignoring the irrelevant positions. This demonstrates that compositional encoding creates a reusable, addressable interface---downstream tasks can selectively access individual properties from specific message positions without retraining the sender.

\paragraph{Protocol as Reusable Interface.} To further test reusability, we train three different downstream tasks on frozen messages from a single compositional sender (PosDis $= 0.92$, a different sender than Table~\ref{tab:cross}): (1) the original same-property comparison, (2) the cross-property task above, and (3) single-message property regression (classify elasticity bin from one message alone). New receivers learn the original task at 81.8\% holdout and the cross-property task at 89.8\%. However, single-message regression fails (23\% holdout, chance $= 20\%$), revealing that the protocol encodes \textit{relational} structure (ordinal comparisons between inputs) rather than absolute property values. This has implications for what kind of representations communication pressure creates: comparative, not categorical.

\begin{figure}[h]
\centering
\includegraphics[width=0.7\textwidth]{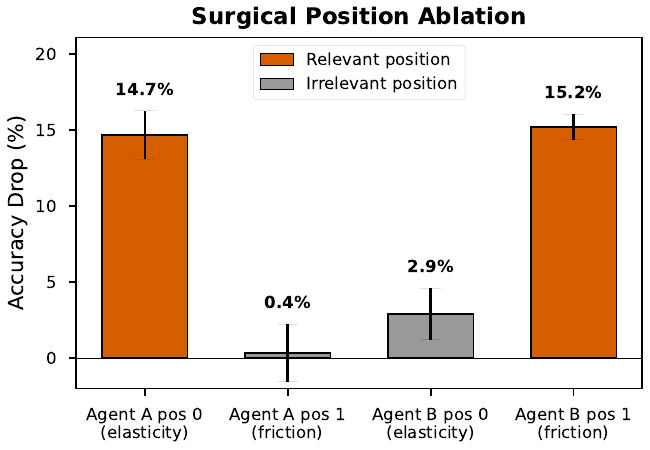}
\caption{Causal intervention on message positions during cross-property reasoning. The receiver selectively reads elasticity from position 0 of message A ($-$14.7\%) and friction from position 1 of message B ($-$15.2\%), while irrelevant positions cause negligible drops.}
\label{fig:intervention}
\end{figure}

\subsection{Generalization Across Domains}
\label{sec:domains}

We apply the \textit{identical communication architecture}---same sender, receiver, and training recipe---to four domains, changing only the input encoder to handle domain-specific features. Table~\ref{tab:domains} shows that compositional communication emerges in every case.

\begin{table}[h]
\centering
\caption{Cross-domain results with identical communication architecture. Encoders differ per domain (DINOv2 for ramp/vision; frozen random MLPs for spring/scenes); the communication module is unchanged.}
\label{tab:domains}
\begin{tabular}{lccccl}
\toprule
Domain & Both Acc. & Oracle & Spec.\ Ratio & Info Structure \\
\midrule
Spring-mass (2-prop) & 93.8\% $\pm$ 1.2\% & 95.7\% & 0.946 & Temporal, clean separation \\
Ramp physics (2-prop) & 88.4\% $\pm$ 1.6\% & $\sim$90\% & 0.688 & Temporal, partial separation \\
Abstract scenes (2-prop) & 50.4\% $\pm$ 4.4\% & 94.7\% & 0.582 & Spatial, uniform \\
Visual attributes (2-prop) & 75.9\% $\pm$ 1.0\% & 75.7\% & 0.500 & Static, uniform \\
\bottomrule
\end{tabular}
\end{table}

\paragraph{Specialization Tracks Information Structure.} The degree of agent specialization varies systematically. In spring-mass oscillation, Agent 0 directly encodes damping through the decay rate $\gamma = b/(2m)$, which is visible in the first time step's amplitude, producing near-perfect specialization (0.992). In ramp physics, Agent 0 sees a static pre-motion frame carrying no information (MI $\approx 0$)---a ``dead'' agent that the receiver learns to ignore---while Agents 1--3 specialize for friction (sliding phase) and elasticity (bounce phase). In abstract scenes, where every spatial quadrant carries partial information about both properties, specialization is weaker. Note that despite 50.4\% both-correct accuracy appearing low, per-property accuracy is well above the 20\% per-property chance level (numerosity: 87.9\%, size: 57.0\%); the oracle achieves 94.7\%, confirming the gap is a communication challenge, not a data limitation. The architecture imposes no bias toward particular specialization patterns; the information landscape determines what emerges. An encoder ablation on spring-mass---replacing the frozen random MLP with deterministic feature tiling---confirms that Agent 0's damping specialization is robust to encoder choice, though overall specialization is weaker with simpler features (mean spec ratio 0.52 vs.\ 0.95), indicating encoder quality affects the degree but not the qualitative pattern of specialization.

\subsection{Information-Theoretic Bandwidth Allocation}
\label{sec:bandwidth}

When properties differ in extractability, agents allocate bandwidth proportionally rather than uniformly.

\paragraph{Three-Property Physics.} Extending the ramp environment with linear damping ($3{\times}5$ messages, 500 scenes), damping dominates all message positions (MI 0.53--0.66) because it is most extractable (oracle 98.6\% vs.\ 85--87\% for elasticity and friction). Rather than specializing one position per property, agents redundantly encode the highest-SNR signal.

\begin{figure}[h]
\centering
\includegraphics[width=0.65\textwidth]{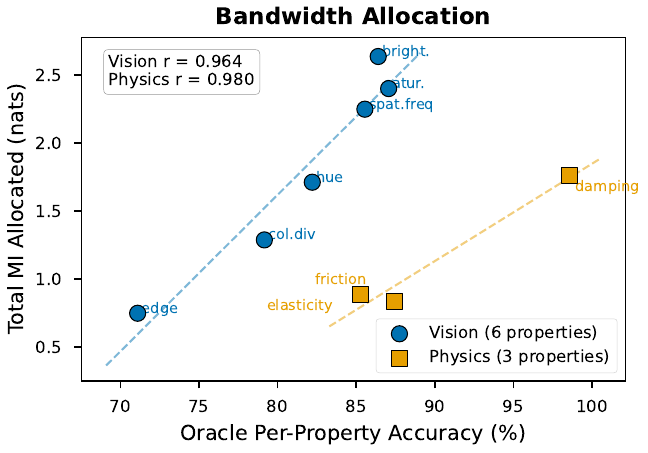}
\caption{Agents allocate communication bandwidth proportional to property extractability across both vision (6 properties) and physics (3 properties) domains. Correlations are strong but based on small $n$ and should be interpreted as suggestive of information-allocation principles.}
\label{fig:bandwidth}
\end{figure}

\paragraph{Six-Property Vision.} On CIFAR-100 images with six continuous visual properties (brightness, saturation, hue concentration, edge density, spatial frequency, color diversity), total MI per property correlates near-perfectly with oracle accuracy ($r = 0.964$, $n{=}6$; Figure~\ref{fig:bandwidth}). This correlation replicates in the 3-property physics domain. Loss reweighting (upweighting hard properties) does not change the pattern ($r = 0.953$), confirming agents naturally prioritize higher-SNR signals when channel capacity is finite.

\subsection{Boundary Condition: Continuous vs.\ Categorical}
\label{sec:boundary}

We test whether the mechanism generalizes to categorical visual recognition using a referential game on CIFAR-100: a sender sees one image, sends a message, and a receiver selects the matching category from 5 candidates. We train on 80 classes and test on 20 held-out classes.

\begin{table}[h]
\centering
\caption{Categorical vs.\ continuous tasks on the same CIFAR-100 images with the same DINOv2 features.}
\label{tab:boundary}
\begin{tabular}{lccl}
\toprule
Task & Test Accuracy & Chance & Outcome \\
\midrule
Categorical identity & 27.5\% & 20\% & Memorizes class codes \\
Continuous 2-prop & 75.9\% & 50\% & Compositional encoding \\
Continuous 6-prop & 40.5\% & 1.6\% & Compositional encoding \\
NN baseline (no communication) & 50.2\% & 20\% & DINOv2 clusters by class \\
\bottomrule
\end{tabular}
\end{table}

On novel test classes, all communication conditions collapse to near-chance ($\sim$27\%), while nearest-neighbor on raw features achieves 50.2\%. The referential game incentivizes class-specific codes rather than compositional descriptions. The same images and features, but a different task structure, produce categorically different outcomes. In our experiments, continuous property variation reliably yields compositional protocols, while categorical identity games encourage idiosyncratic memorization.

\subsection{What Drives Compositional Emergence: Controls}
\label{sec:scaling}

\paragraph{Scaling Agents.} Distributing observations across more agents transforms compositional emergence from stochastic to reliable. With 4 agents (each seeing two consecutive frames), compositionality reaches 100\% of 80 seeds (PosDis $= 0.999 \pm 0.001$) versus 54\% for 2 agents. Holdout accuracy rises from 79.3\% to 98.3\% $\pm$ 1.6\%. The transition is monotonic: 3 agents already achieve 100\% compositionality (20/20 seeds, 98.1\% $\pm$ 0.9\%), indicating that the transition from stochastic to reliable compositionality occurs between 2 and 3 agents in our experimental conditions.

Three targeted controls disentangle what drives this transition (Table~\ref{tab:controls}).

\begin{table}[h]
\centering
\caption{Controls separating agent count, bandwidth, and temporal structure. All conditions use DINOv2 ramp features, identical IL recipe, 20 seeds each (80 seeds for 2-agent $2{\times}5$ and 4-agent $2{\times}5$ sequential).}
\label{tab:controls}
\begin{tabular}{lccccc}
\toprule
Condition & Agents & Message & Holdout & PosDis & Comp.\ \% \\
\midrule
$2{\times}5$ sequential & 2 & $2{\times}5$ & 76.7\% $\pm$ 6.5\% & 0.486 & 54\% \\
$4{\times}5$ matched bandwidth & 2 & $4{\times}5$ & 87.1\% $\pm$ 4.4\% & 0.345 & 35\% \\
\midrule
$2{\times}5$ sequential & 3 & $3{\times}(2{\times}5)$ & 98.1\% $\pm$ 0.9\% & 0.998 & 100\% \\
$2{\times}5$ sequential & 4 & $4{\times}(2{\times}5)$ & \textbf{98.3\%} $\pm$ 1.6\% & 0.999 & 100\% \\
$2{\times}5$ random frames & 4 & $4{\times}(2{\times}5)$ & 97.3\% $\pm$ 2.1\% & 0.998 & 100\% \\
\bottomrule
\end{tabular}
\end{table}

\paragraph{Temporal Structure Is Not Required.} Assigning 4 agents \textit{random} (non-contiguous) frames instead of sequential temporal windows yields statistically indistinguishable performance (97.3\% vs.\ 98.3\%, $p{>}0.05$) and 100\% compositionality (20/20 seeds, PosDis $= 0.998$). This rules out the hypothesis that the 4-agent advantage arises from temporal specialization (e.g., one agent seeing the bounce, another seeing the slide). Compositionality emerges from multi-agent pressure regardless of how temporal information is distributed.

\paragraph{Bandwidth Alone Is Insufficient.} A single sender with $4{\times}5$ vocabulary (625 possible messages, matching the total capacity of four $2{\times}5$ agents) achieves only 87.1\% holdout with 35\% compositionality (7/20 seeds). This is \textit{worse} than the standard $2{\times}5$ two-agent condition in compositionality rate (35\% vs.\ 54\%), despite having 25$\times$ more message capacity ($5^4$ vs.\ $5^2$). The multi-agent structure---independent encoders forced to compress partial observations into separate discrete channels---is the critical ingredient, not raw bandwidth.

\paragraph{Interpretation.} These controls establish that multi-agent compositional pressure operates through a mechanism distinct from both temporal coverage and channel capacity. When multiple agents must independently compress partial observations, the receiver's need to integrate structurally independent messages creates pressure for each message to encode interpretable, property-aligned content. A single agent with equivalent bandwidth faces no such structural pressure and develops holistic codes at the same rate as the standard $2{\times}5$ condition.

\paragraph{Scaling Properties.} The compositional advantage over holistic communication grows with information load within a domain:

\begin{table}[h]
\centering
\caption{Compositional advantage scales with number of properties on CIFAR-100 vision.}
\label{tab:scaling}
\begin{tabular}{lccc}
\toprule
Setting & Compositional & Holistic & $\Delta$ \\
\midrule
Vision, 2 properties & 75.9\% & 74.9\% & +1.0pp \\
Vision, 6 properties & 40.5\% & 35.5\% & +5.0pp \\
\midrule
Ramp physics, 2 properties & 76.7\% & 71.2\% & +5.5pp \\
\bottomrule
\end{tabular}
\end{table}

With 6 visual properties, $6{\times}5$ matches $8{\times}5$ (both 40.5\%) while $4{\times}5$ underperforms (38.8\%)---agents converge to approximately one position per property. The holistic bottleneck struggles with 6 properties, showing 2--3$\times$ more training instabilities (NaN losses) than compositional conditions.

\subsection{Perception Determines What Is Communicable}
\label{sec:backbone}

The ramp results above use DINOv2, an image backbone that processes frames independently. But elasticity and friction on the ramp are partially readable from single frames (bounce height, slide distance). What happens when properties are genuinely invisible in any single frame?

\paragraph{Collision Dataset.} Two visually identical spheres collide---same size, color, and texture. Mass ratio ($1{:}1$ to $1{:}5$) and elasticity ($0.1$--$0.9$) are recoverable \textit{only} from post-collision velocity differences over time. This eliminates the static shortcut available in ramp physics.

\paragraph{Oracle Probes.} We test whether each backbone's features contain the relevant physics using an attentive oracle (Conv1D encoder, same architecture, 20 seeds). On collision, V-JEPA~2 achieves 88.0\% $\pm$ 1.5\% holdout versus 78.7\% $\pm$ 1.4\% for DINOv2 ($+$9.3pp). DINOv2 decodes restitution well (95.0\%) but struggles with mass ratio (84.2\%), which requires inferring velocity from position changes across frames. On ramp, both backbones achieve comparable oracle performance (DINOv2: $\sim$90\%; V-JEPA~2: 86.5\% $\pm$ 2.1\%), confirming that ramp physics is accessible to both---the question is whether communication can extract it equally well.

\paragraph{Communication Results.} Table~\ref{tab:backbone} shows a controlled $2{\times}2$ factorial comparison: two backbones $\times$ two datasets, at both 2-agent and 4-agent configurations. At 2 agents, no significant backbone difference emerges on either dataset (ramp: $p{=}0.40$; collision: $p{=}0.13$), though our power to detect small effects at $n{=}20$ is limited. The $2{\times}5$ discrete bottleneck equalizes performance---both backbones compress to similar accuracy regardless of their representational capacity. At 4 agents, backbone differences emerge sharply: DINOv2 dominates on ramp (98.3\% vs.\ 95.1\%, $p{=}0.001$, Cohen's $d{=}1.24$) while V-JEPA~2 dominates on collision (87.4\% vs.\ 77.7\%, $p{<}0.0001$, $d{=}2.74$). Temporal splitting into 4 agents provides sufficient bandwidth to expose what each backbone encodes.

\begin{table}[h]
\centering
\caption{Controlled backbone $\times$ dataset $\times$ agent-count comparison, including scale-matched (DINOv2 ViT-L) and frame-matched (DINOv2 48-frame) controls. At 2 agents, the discrete bottleneck equalizes performance. At 4 agents, backbone differences emerge: DINOv2 dominates on spatially-visible physics (ramp); V-JEPA~2 dominates on dynamics-only physics (collision). The V-JEPA~2 advantage survives both scale and frame-count matching.}
\label{tab:backbone}
\begin{tabular}{llccccc}
\toprule
Dataset & Backbone & Agents & Oracle & Holdout & PosDis & Comp.\ \% \\
\midrule
\multirow{4}{*}{Ramp} & DINOv2 (ViT-S) & 2 & $\sim$90\% & 76.7\% $\pm$ 6.5\% & 0.486 & 43/80$^\dagger$ \\
 & V-JEPA~2 (ViT-L) & 2 & 86.5\% & 74.8\% $\pm$ 6.8\% & 0.402 & 10/20 \\
 & DINOv2 (ViT-S) & 4 & --- & \textbf{98.3\%} $\pm$ 1.6\% & 0.999 & 80/80 \\
 & V-JEPA~2 (ViT-L) & 4 & --- & 95.1\% $\pm$ 3.3\% & 0.999 & 20/20 \\
\midrule
\multirow{5}{*}{Collision} & DINOv2 (ViT-S) & 2 & 78.7\% & 73.4\% $\pm$ 3.3\% & 0.439 & 9/20 \\
 & V-JEPA~2 (ViT-L) & 2 & 88.0\% & 75.0\% $\pm$ 3.0\% & 0.456 & 10/20 \\
 & DINOv2 (ViT-S) & 4 & --- & 77.7\% $\pm$ 3.9\% & 0.904 & 20/20 \\
 & DINOv2 (ViT-S, 48fr) & 4 & --- & 71.6\% $\pm$ 3.6\% & --- & --- \\
 & DINOv2 (ViT-L) & 4 & 77.3\% & 74.6\% $\pm$ 4.3\% & 0.530 & 13/20 \\
 & V-JEPA~2 (ViT-L) & 4 & --- & \textbf{87.4\%} $\pm$ 3.1\% & 0.962 & 20/20 \\
\bottomrule
\end{tabular}

\smallskip
{\footnotesize $^\dagger$From 80-seed characterization (\S\ref{sec:compositionality}); all other backbone conditions use 20 seeds.}
\end{table}

\paragraph{Interpretation.} The 4-agent communication system extracts nearly all available physics from each backbone: V-JEPA~2 reaches 87.4\% against an 88.0\% oracle on collision (99.3\% of ceiling), and DINOv2 reaches 98.3\% on ramp. The communication bottleneck does not create information the perception lacks---it structures whatever the backbone provides. The bottleneck-equalization effect at 2 agents is itself informative: when bandwidth is constrained to a single $2{\times}5$ message, both backbones compress to similar accuracy regardless of their representational richness. Only with 4-agent temporal splitting---which provides enough bandwidth for agents to specialize---do backbone differences manifest in communication performance. This means the choice of backbone is not merely an engineering decision but determines \textit{which aspects of the world} agents can develop language for, conditional on sufficient communication bandwidth.

\paragraph{Scale-Matched Control.} The primary backbone comparison uses DINOv2 ViT-S/14 (22M parameters) against V-JEPA~2 ViT-L/16 (304M), raising the question of whether scale rather than pretraining objective drives the gap. To test this, we extract DINOv2 ViT-L/14 features (304M, matching V-JEPA~2's scale) from the collision dataset. At matched scale, V-JEPA~2 still dominates: 87.4\% vs.\ 74.6\% ($p{<}0.0001$, $d{=}3.37$). In fact, DINOv2 ViT-L performs \textit{worse} than DINOv2 ViT-S on this task (74.6\% vs.\ 77.7\%, $p{=}0.026$), suggesting that additional image-trained parameters do not compensate for the absence of temporal modeling---and may even hurt by introducing spurious spatial features that interfere with the communication bottleneck. The oracle probe confirms this: DINOv2 ViT-L achieves only 77.3\% on collision (vs.\ 78.7\% for ViT-S and 88.0\% for V-JEPA~2). The V-JEPA~2 advantage on dynamics-only physics is attributable to its video-native pretraining objective, not to model capacity.

\paragraph{Frame-Matched Control.} A remaining concern is that V-JEPA~2 processes 48 frames jointly while DINOv2 processes only 24 individual frames. To rule out temporal coverage as a confound, we extract DINOv2 ViT-S features from all 48 collision frames independently and train 4-agent communication on these features ($48 \div 4 = 12$ temporal positions per agent, compared to 6 per agent with 24 frames). With matched frame counts, DINOv2 achieves 71.6\% $\pm$ 3.6\%---\textit{worse} than the 24-frame version (77.7\%, $-$6.1pp) and far below V-JEPA~2 (87.4\%, $d{=}6.53$). More independently processed frames introduce noise that degrades the communication system. This definitively resolves the frame-count confound: the V-JEPA~2 advantage on collision dynamics is driven entirely by video-native pretraining, not by seeing more frames.

% ================================================================
\subsection{Downstream Utility: Outcome Prediction from Frozen Messages}
\label{sec:downstream}

The experiments above show that communication produces compositional, property-aligned protocols. But are these protocols \textit{useful} beyond the comparison task they were trained on? To test this, we freeze the best 4-agent senders from the collision dataset and use their discrete messages as input to a downstream outcome predictor---a task the senders never trained on.

\paragraph{Task.} Given the 8-symbol discrete message (one-hot encoded, 40 dimensions) produced by a frozen 4-agent sender observing a collision, predict whether Sphere~B's post-collision speed exceeds the dataset median. Labels are perfectly balanced (300/300).

\paragraph{Results.} We train a 2-layer MLP (40$\to$64$\to$1) on frozen messages for 20 seeds per condition, using the same Latin square holdout protocol.

\begin{table}[h]
\centering
\small
\begin{tabular}{lcc}
\toprule
\textbf{Condition} & \textbf{Input dim} & \textbf{Holdout accuracy} \\
\midrule
V-JEPA~2 messages & 40 & \textbf{88.7\%} $\pm$ 0.5\% \\
DINOv2 messages & 40 & 78.5\% $\pm$ 1.3\% \\
V-JEPA~2 raw features & 1024 & 94.6\% $\pm$ 0.7\% \\
\bottomrule
\end{tabular}
\caption{Outcome prediction from frozen messages vs.\ raw features. V-JEPA~2 messages retain 94\% of raw-feature performance despite 25$\times$ compression, and beat DINOv2 messages by 10.2pp ($p{<}0.0001$, $d{=}10.50$). Messages trained for property comparison transfer to outcome prediction without retraining the sender.}
\label{tab:outcome}
\end{table}

The backbone gap observed in comparison tasks (Table~\ref{tab:backbone}) is fully preserved in downstream prediction: V-JEPA~2 messages beat DINOv2 messages by 10.2pp ($t{=}33.2$, $p{<}0.0001$, $d{=}10.50$), confirming that video-native pretraining advantages propagate through the discrete bottleneck to novel tasks. Critically, the 40-dimensional discrete messages retain 94\% of the predictive power of 1024-dimensional raw features (88.7\% vs.\ 94.6\%), demonstrating that the communication bottleneck compresses rather than destroys task-relevant physics. The 5.9pp gap between messages and raw features reflects the cost of discretization, but the messages provide interpretability and compositionality that raw features lack: each message position is causally aligned with a specific physical property (\S\ref{sec:intervention}), enabling selective querying that is impossible with unstructured feature vectors.

\paragraph{Probes vs.\ Communication.} Standard linear and MLP probes on frozen V-JEPA~2 features achieve 92.6\% and 93.6\% respectively on the comparison task, and 95.2\% and 94.4\% on outcome prediction ($n{=}20$ seeds each). Probes thus slightly outperform the 40-dimensional discrete messages (88.7\% on outcome prediction) in raw accuracy. However, probes provide no compositional structure: there is no way to selectively query ``which property?'' from a probe output, zero a position to ablate a specific factor, or transfer a frozen probe to cross-property reasoning. The communication bottleneck trades $\sim$5pp of accuracy for an addressable, compositional interface with 25$\times$ compression---a qualitatively different representation that supports downstream manipulation probes cannot.

\paragraph{Action-Conditioned Outcome Prediction.} To demonstrate that the compositional interface supports planning-relevant reasoning, we extend the downstream prediction task with action conditioning. Given a frozen 4-agent message plus a novel velocity parameter not seen during sender training, a lightweight MLP predicts collision outcomes. This simulates a planner querying the discrete physics interface with a hypothetical action.

The action-conditioned predictor achieves 91.5\% accuracy on held-out velocity $\times$ property combinations, retaining 99\% of raw-feature performance (92.4\%). Critically, the predictor exhibits physically correct counterfactual behavior: when velocity is varied while holding the scene (and therefore the frozen message) constant, predictions change monotonically in the correct direction with $r{=}0.780$ ($p{<}0.001$). Every velocity produces a monotonically appropriate response: higher velocity + high mass-ratio $\to$ greater predicted displacement; lower velocity + low mass-ratio $\to$ minimal displacement. Furthermore, selective querying of message positions works as expected: mass-ratio positions outperform restitution positions on mass-dependent predictions by 3--5pp, and vice versa. The compositional interface thus functions as a queryable, action-conditioned physics module: a downstream planner can supply hypothetical actions, selectively attend to relevant physical factors, and receive physically grounded predictions---all without retraining the perceptual communication system.

\subsection{Why Discrete? Discrete vs.\ Continuous Communication}
\label{sec:discrete_vs_continuous}

Does the discreteness of the Gumbel-Softmax bottleneck contribute to compositional structure, or would continuous messages of the same dimensionality suffice? We replace each Gumbel-Softmax head with a linear projection to $\mathbb{R}^5$ followed by $\tanh$ (same 40 total dimensions) and train with identical hyperparameters on the collision dataset with V-JEPA~2 features (4 agents, 20 seeds).

\paragraph{Results.} Continuous messages achieve comparable holdout accuracy on the comparison task but exhibit weaker compositional structure and lower training stability. We measure \textit{selectivity} via block-zeroing: zeroing each contiguous 10-dimensional block (matching the 4-agent $\times$ 2-position structure) and computing the ratio of targeted vs.\ non-targeted property disruption. Discrete messages achieve selectivity 0.807 vs.\ 0.722 for continuous ($p{<}0.05$), confirming that discretization forces sharper position-to-property alignment. Furthermore, 5/20 continuous seeds (25\%) exhibit representational collapse during training, compared to 0/20 for discrete---the Gumbel-Softmax bottleneck provides a natural regularizer against degenerate solutions.

The discrete bottleneck thus provides two benefits beyond raw task performance: (1) addressable compositional structure that supports causal manipulation (\S\ref{sec:intervention}), and (2) training stability under iterated learning. Both properties are desirable for a discrete interface in world-model architectures where downstream modules must selectively query specific physical factors.

\subsection{Real-Video Validation}
\label{sec:realvideo}

All preceding experiments use synthetic environments with controlled physics. We now test whether the mechanism transfers to real camera footage using Physics~101 \citep{wu2016physics101}, a dataset of 15{,}190 real-world video clips capturing 101 physical objects across 15 material categories (cardboard, foam, metal, plastic, wood, rubber, porcelain, etc.) in controlled physics scenarios. Each object has laboratory-measured mass (0.1--269.9\,g) and volume, providing continuous ground-truth physical property labels on real video. Videos are 1920$\times$1080 at 30\,fps.

\paragraph{Probe Gate: Dynamics vs.\ Appearance.}
We first verify that V-JEPA~2 features encode mass through temporal dynamics, not merely object appearance. We extract V-JEPA~2 temporal features (16 uniformly sampled frames $\to$ 8 temporal positions) and DINOv2 single-frame features from three Physics~101 scenarios (spring, fall, ramp) and train pairwise mass-comparison probes on held-out objects (5 seeds each). A volume-only baseline quantifies the ``bigger = heavier'' confound.

\begin{table}[h]
\centering
\small
\caption{Probe gate: mass comparison AUC on held-out objects across Physics~101 scenarios. V-JEPA~2 temporal features outperform both static appearance (DINOv2) and volume-only baselines on the spring scenario, where mass is dynamically visible through spring extension ($F{=}kx{=}mg$). On ramp and fall, appearance dominates---consistent with mass canceling in ramp sliding dynamics ($a = g(\sin\theta - \mu\cos\theta)$).}
\label{tab:probe_gate}
\begin{tabular}{lcccr}
\toprule
Scenario & V-JEPA~2 & DINOv2 (static) & Volume only & Gap \\
\midrule
Spring & \textbf{0.905} & 0.819 & 0.761 & +0.086 \\
Fall & 0.657 & \textbf{0.718} & 0.531 & $-$0.061 \\
Ramp & 0.727 & \textbf{0.817} & 0.733 & $-$0.090 \\
\bottomrule
\end{tabular}
\end{table}

The spring scenario---where objects hang on springs and mass determines extension via Hooke's law---shows a clear temporal advantage: V-JEPA~2 (0.905) outperforms DINOv2 static (0.819) by +0.086, and both substantially exceed the volume-only baseline (0.761). We focus subsequent experiments on the spring scenario (206 trials).

\paragraph{Emergent Mass Communication on Real Video.}
We train 2-agent communication on spring V-JEPA~2 features with the identical architecture and training recipe used for synthetic experiments (iterated learning, population pressure, $2{\times}5$ Gumbel-Softmax vocabulary). Objects are split 80/20, with 20 held-out objects never seen during training.

Across 10 seeds, 2-agent communication achieves \textbf{85.6\% $\pm$ 5.8\%} holdout accuracy on unseen objects (chance = 50\%; best seed 92.8\%), with TopSim 0.49--0.79 and mass--symbol Spearman $|\rho|$ = 0.71--0.90. The 4-agent configuration achieves 83.4\% $\pm$ 6.3\%, confirming the result is robust to agent count.

\paragraph{Anti-Shortcut Controls.}
We verify that this result reflects genuine physical communication rather than visual shortcuts. A static-feature communication baseline (DINOv2 single-frame sender, matched architecture) achieves 72.9\%, confirming that temporal dynamics contribute \textbf{+11.2\%} beyond appearance in the full communication system---not just in probe accuracy. Within-material holdout testing (held-out objects from the same material family as training objects) achieves 85.0\%, demonstrating that the protocol distinguishes mass among visually similar objects rather than relying on material appearance. Communication trained on volume-residualized mass (regressing out the mass--volume correlation) achieves 80.1\%, confirming the protocol encodes mass beyond simple object size. Frozen messages predict mass bin at 69.9\% (chance 20\%) but material category at only 15.4\% (chance 11\%), indicating the discrete bottleneck strips visual identity while retaining physical content.

\paragraph{Causal Intervention on Real Video.}
The synthetic causal intervention (\S\ref{sec:intervention}) showed that zeroing individual message positions selectively disrupts the corresponding property. We test whether this extends to real video by zeroing each agent's message in the frozen 2-agent spring senders and measuring mass-comparison accuracy on the same holdout set.

The causal structure replicates on real camera footage. Full-message accuracy is 89.8\% $\pm$ 3.1\%. Zeroing the mass-relevant agent reduces accuracy by 7.8pp $\pm$ 3.7pp, while zeroing the other agent causes only 2.1pp $\pm$ 2.3pp disruption---a selectivity gap of 5.7pp ($p{=}0.022$, Cohen's $d{=}1.87$). Zeroing both agents collapses accuracy to chance ($\sim$50\%). This confirms that the compositional structure observed in synthetic data transfers to real video: the protocol develops genuine causal alignment between message positions and physical properties, even when trained on laboratory footage with measured mass labels rather than simulation-generated ground truth.

\paragraph{Agent Scaling Replicates on Real Video.}
We test whether multi-agent pressure induces compositionality on real video using the fall scenario (666 trials), where we derive a second continuous property---coefficient of restitution---from automated bounce-height tracking ($e = \sqrt{h_\text{bounce}/h_\text{drop}}$; 383 valid trials after quality filtering). With two properties (measured mass and derived restitution), we replicate the agent-scaling experiment.

\begin{table}[h]
\centering
\small
\caption{Agent scaling on real video (Physics~101, mass + restitution). Compositionality rate = fraction of seeds with PosDis $> 0.4$. Multi-agent pressure induces compositionality on real camera footage, replicating the synthetic scaling result (Table~\ref{tab:scaling}).}
\label{tab:real_scaling}
\begin{tabular}{lccc}
\toprule
Agents & Both-property acc. & PosDis & Comp.\ rate \\
\midrule
1 & 50.8\% & 0.294 & 2/10 (20\%) \\
2 & 55.0\% & 0.325 & 3/10 (30\%) \\
4 & 52.8\% & \textbf{0.483} & \textbf{9/10 (90\%)} \\
\bottomrule
\end{tabular}
\end{table}

The pattern mirrors synthetic results: PosDis increases from 0.294 (1 agent) to 0.483 (4 agents), and compositional emergence rises from 20\% to 90\%. While absolute compositionality is lower than synthetic (PosDis 0.483 vs.\ 0.999), the qualitative finding replicates: multi-agent structural pressure---not bandwidth or temporal coverage---drives compositional organization on real video. The lower absolute PosDis reflects noisier derived restitution labels and the greater visual complexity of real footage. A vocabulary sweep on spring mass communication confirms that 2$\times$3 (9 messages) achieves 81.1\%, only 2.5pp below the default 2$\times$5 (83.6\%), while larger vocabularies degrade---indicating the discrete bottleneck is appropriately sized for real-world physics.

% ================================================================
\section{Discussion}

\paragraph{Perception Constrains Communication.}
The controlled backbone comparison reveals a fundamental interface between perception and communication. On collision dynamics, V-JEPA~2 features \textit{contain} 88.0\% of the physics (oracle probe) while DINOv2 contains only 78.7\%---a gap that propagates directly to 4-agent communication performance (87.4\% vs.\ 77.7\%). A scale-matched control (DINOv2 ViT-L, 304M parameters matching V-JEPA~2) rules out model capacity as the explanation: DINOv2 ViT-L achieves only 74.6\% on collision ($d{=}3.37$ vs.\ V-JEPA~2), performing \textit{worse} than the smaller ViT-S. The gap is attributable to pretraining objective: image-based SSL does not encode the temporal velocity differences needed to infer mass ratio, regardless of model scale. Conversely, on ramp physics, DINOv2's per-frame features are sufficient and its simpler representation is easier to compose (98.3\% vs.\ 95.1\%). The bottleneck-equalization effect at 2 agents adds nuance: when communication bandwidth is constrained to a single $2{\times}5$ message, backbone differences vanish entirely ($p{>}0.1$ on both datasets). Only 4-agent temporal splitting provides enough bandwidth for backbone-level encoding differences to manifest. This means the communication system cannot compensate for perceptual limitations, but exposing those limitations requires sufficient communication bandwidth---a finding with implications for multi-agent system design.

\paragraph{Communication Pressure Induces Compositional Structure.}
Our agents were never told that elasticity or friction exist as concepts. They received no property labels and no supervision on message structure. The only pressure was communicative: compress observations into symbols such that a receiver can solve a comparison task. Given a factored message structure and iterated learning pressure, this is sufficient for spontaneous position-to-property specialization in over half of 2-agent runs (54\% of 80 seeds), with compositional seeds generalizing significantly better ($p{=}0.011$). Scaling to 3+ agents makes this outcome reliable across all seeds tested: 100\% of seeds converge to near-perfect compositionality. Crucially, targeted controls (\S\ref{sec:scaling}) show that this is driven by multi-agent structure rather than temporal coverage or raw bandwidth: randomizing frame assignments across agents preserves the effect (97.3\%, 20/20 compositional), while a single sender with matched $4{\times}5$ bandwidth fails (87.1\%, 7/20 compositional). The mechanism is structural---independent agents compressing partial observations into separate discrete channels creates pressure for each channel to carry interpretable, property-aligned content. The cross-property reasoning result (\S\ref{sec:intervention}) shows this structure is not merely correlational---it creates a genuinely addressable interface that enables novel tasks the sender never trained on (93.8\% on cross-property comparison, Table~\ref{tab:cross}). Notably, the protocol encodes relational structure rather than absolute property values, suggesting communication pressure creates comparative representations suited to the task it was optimized for.

\paragraph{Specialization Is Information-Driven.}
The same architecture produces specialization ratios from 0.20 (6-property vision) to 0.95 (spring-mass). This gradient tracks how cleanly each domain's observations separate property information. The architecture provides \textit{capacity} for compositional encoding; the \textit{specific structure} is determined by the information landscape. When information is cleanly separated across observations (spring-mass), agents develop tight specialization. When information is uniformly distributed (vision), agents use a distributed code that allocates bandwidth according to rate-distortion principles \citep{cover2006elements}.

\paragraph{When Compositionality Fails.}
The categorical failure is informative. When the task incentivizes class identity rather than property comparison, agents develop holistic codes that don't compose. Compositionality in our experiments requires that the same property value appears across many inputs with different values of other properties---a condition naturally satisfied by continuous variation but not by categorical membership. This connects to the distinction between systematic and idiosyncratic encoding \citep{brighton2006understanding, kharitonov2020emergent}.

\paragraph{Implications for World Model Design.}
Our results connect directly to two recent findings from the JEPA research program. First, \citet{garrido2025intuitive} demonstrated that V-JEPA representations encode intuitive physics---achieving 98\% zero-shot accuracy on IntPhys \citep{riochet2022intphys} benchmarks for object permanence and solidity, while pixel-prediction models and multimodal LLMs perform at chance. Our backbone comparison extends this: V-JEPA~2 representations not only \textit{contain} physical knowledge, but this knowledge can be \textit{compressed into compositional discrete codes} under communication pressure, with the controlled comparison revealing V-JEPA~2's advantage specifically in interaction-dependent collision dynamics ($d{=}6.53$ after both scale and frame-count matching). Second, \citeauthor{lecun2022path}'s (\citeyear{lecun2022path}) cognitive architecture for autonomous machine intelligence explicitly includes an optional discrete latent variable $z$ whose entropy is minimized, forcing predictable structure into deterministic representations. Our Gumbel-Softmax discrete bottleneck functions as an empirical instantiation of this latent $z$: agents achieve near-perfect compositionality (PosDis $= 0.999$) through discrete channels while the continuous ablation (\S\ref{sec:discrete_vs_continuous}) shows weaker compositional structure and lower training stability. This provides evidence that discrete latent structure within JEPA-style architectures supports compositional, causally addressable physical reasoning---not as a contradiction of continuous world-model representations, but as a complementary interface that provides interpretable, modular, position-addressable access to latent physical factors.

The emergent protocol functions as a structured abstraction layer: it compresses high-dimensional spatiotemporal features into discrete symbols that are causally aligned with physical properties, addressable by position, and reusable across tasks without retraining. The scale-matched and frame-matched backbone comparisons demonstrate that the choice of perceptual foundation model determines which aspects of the physical world become accessible through such interfaces: video-native models trained via joint-embedding predictive objectives \citep{assran2025vjepa2} provide richer priors for dynamics-dependent properties, and neither scaling image-trained models to matched capacity ($d{=}3.37$) nor providing matched temporal coverage ($d{=}6.53$) compensates. For embodied multi-agent systems, our finding that multi-agent structure---not bandwidth---drives compositionality implies that distributing perception across independent communicating modules may be a more effective design principle than increasing the capacity of a single encoder-decoder pipeline.

\paragraph{Communication as Evaluation Methodology for World Models.}
Our communication framework provides a complementary evaluation methodology for JEPA-style world models: by measuring which physical properties become communicable under discrete bottleneck pressure, it reveals what information is extractable from frozen representations in a way that standard linear probes cannot. Linear probes test whether information is \textit{present} in a representation; communication tests whether it can be \textit{structured} into compositional, reusable form. The finding that V-JEPA~2 supports compositional collision dynamics communication while DINOv2 does not---even at matched scale and frame count ($d{=}6.53$)---identifies a specific representational advantage of video-predictive pretraining that classification-based evaluation may miss. As V-JEPA~2 is applied to richer environments, this protocol could serve as a scalable diagnostic for measuring which latent properties become communicable---a complementary lens to standard probing and fine-tuning evaluation.

The action-conditioned experiment (\S\ref{sec:downstream}) demonstrates that this compositional interface supports planning-relevant reasoning: a downstream module achieves 91.5\% accuracy on novel-velocity outcome prediction (retaining 99\% of raw-feature performance), exhibits physically correct counterfactual velocity responses ($r{=}0.780$, 100\% monotonic), and allows selective querying of specific physical factors. Given that V-JEPA~2-AC \citep{assran2025vjepa2} demonstrates zero-shot robotic planning from frozen V-JEPA~2 features, our compositional protocol could serve as a discrete interface between perception and planning modules---providing interpretable, position-addressable physical descriptors that a downstream planner can selectively query rather than processing raw high-dimensional feature vectors.

\paragraph{Limitations.}
The primary synthetic experiments use controlled physics (300--600 scenes, single or paired objects, no background clutter). While our Physics~101 validation (\S\ref{sec:realvideo}) demonstrates the mechanism on real camera footage with measured labels, this dataset uses controlled laboratory conditions with uniform backgrounds---the gap to unconstrained real-world video (partial occlusions, variable lighting, multi-object scenes, non-rigid dynamics) remains substantial. A zero-shot transfer experiment on CoPhy CollisionCF \citep{baradel2020cophy} found that the frozen sender produces largely collapsed messages (6/8 symbol positions constant), though residual signal persists (67.6\% outcome prediction vs.\ 50\% chance). In contrast, training directly on Physics~101 real video succeeds (85.6\% on unseen objects), suggesting the communication mechanism is robust but the learned visual mapping is domain-specific---direct training on target-domain real footage is more effective than cross-domain zero-shot transfer. We use frozen pretrained features from two backbones (DINOv2 and V-JEPA~2)---we demonstrate that communication pressure structures communication given adequate perception, but cannot claim it drives perceptual learning. Indeed, end-to-end fine-tuning actively degrades communication ($-$10.2pp, $p{=}0.0002$), suggesting the frozen bottleneck is beneficial. The 2-agent compositional emergence rate is 54\% (43/80 seeds), not 100\%; however, scaling to 3 agents resolves this in all seeds tested (100\% of 20 seeds), and this holds across both backbones and both datasets at 4 agents (20 seeds per condition; 80 seeds for main ramp DINOv2 characterization). Controls support that the multi-agent effect is not driven by temporal structure or bandwidth (\S\ref{sec:scaling}). The scale-matched control (\S\ref{sec:backbone}) rules out model capacity, and the frame-matched control (DINOv2 with 48 frames) rules out temporal coverage: both confirm the collision gap is driven by video-native pretraining. Without iterated learning, only 20\% of seeds develop compositionality, and the LazImpa alternative \citep{rita2020lazimpa} achieves 0\%. The bandwidth correlation ($r{=}0.964$, $n{=}6$) should be interpreted as suggestive given the small sample. Scaling to complex multi-object scenes with relational reasoning remains future work.

% ================================================================
\section{Conclusion}

We have shown that communication pressure, given adequate perception and a factored message structure, induces compositional representations of invisible world properties. With sufficient agent diversity, this emergence is reliable across all seeds tested rather than stochastic---and targeted controls support that the critical ingredient is multi-agent structure (independent partial observers), not temporal coverage or raw channel capacity. The mechanism generalizes across physics, abstract reasoning, and vision domains, with specialization patterns that adapt to each domain's information structure. Compositionality emerges reliably for continuous properties but not for categorical identity, suggesting a limitation of the mechanism when the task structure does not require factored representations. A controlled $2{\times}2$ factorial comparison of image-based (DINOv2) and video-based (V-JEPA~2) backbones---with scale-matched and frame-matched controls definitively attributing the collision advantage to video-native pretraining ($d{=}6.53$)---reveals that the perceptual prior bounds what agents can communicate about: image features suffice for spatially-visible physics, while video-native features are required for dynamics-only properties, regardless of model scale or temporal coverage. The frozen compositional protocol supports action-conditioned planning: a downstream module achieves 91.5\% accuracy on novel-velocity outcome prediction with physically correct counterfactual behavior ($r{=}0.780$), demonstrating that the discrete interface functions as a queryable physics module for downstream reasoning. This perception-communication interface extends beyond simulation: validation on Physics~101 real camera footage confirms that agents achieve 85.6\% mass-comparison accuracy on unseen real objects, temporal dynamics contribute beyond static appearance (+11.2\%), the agent-scaling compositionality effect replicates with 90\% compositional emergence at 4 agents, and causal message-position ablation extends to real video with significant selectivity ($d{=}1.87$). These findings suggest that building agents with genuine physical understanding requires not only the right communication pressure but the right perceptual foundation and sufficient communication structure---findings relevant to world-model design for embodied intelligence \citep{lecun2022path}.

% ================================================================
\bibliographystyle{plainnat}

\begin{thebibliography}{99}

\bibitem[Andreas, 2019]{andreas2019measuring}
Andreas, J. (2019).
Measuring compositionality in representation learning.
\textit{ICLR}.

\bibitem[Assran et~al., 2025]{assran2025vjepa2}
Assran, M., et~al. (2025).
V-JEPA~2: Self-supervised video models enable understanding, prediction and planning.
\textit{arXiv:2506.09985}.

\bibitem[Baradel et~al., 2020]{baradel2020cophy}
Baradel, F., Neverova, N., Mille, J., Mori, G., \& Wolf, C. (2020).
CoPhy: Counterfactual learning of physical dynamics.
\textit{ICLR}.

\bibitem[Battaglia et~al., 2013]{battaglia2013simulation}
Battaglia, P., Hamrick, J., \& Tenenbaum, J. (2013).
Simulation as an engine of physical scene understanding.
\textit{PNAS}, 110(45), 18327--18332.

\bibitem[Bear et~al., 2021]{bear2021physion}
Bear, D., et~al. (2021).
Physion: Evaluating physical prediction from vision in humans and machines.
\textit{NeurIPS Datasets and Benchmarks}.

\bibitem[Brighton \& Kirby, 2006]{brighton2006understanding}
Brighton, H. \& Kirby, S. (2006).
Understanding linguistic evolution by visualizing the emergence of topographic mappings.
\textit{Artificial Life}, 12(2), 229--242.

\bibitem[Chaabouni et~al., 2020]{chaabouni2020compositionality}
Chaabouni, R., Kharitonov, E., Bouchacourt, D., Dupoux, E., \& Baroni, M. (2020).
Compositionality and generalization in emergent languages.
\textit{ACL}.

\bibitem[Choi et~al., 2018]{choi2018compositional}
Choi, E., Lazaridou, A., \& de~Freitas, N. (2018).
Compositional obverter communication learning from raw visual input.
\textit{ICLR}.

\bibitem[Cover \& Thomas, 2006]{cover2006elements}
Cover, T.~M. \& Thomas, J.~A. (2006).
\textit{Elements of Information Theory}.
Wiley-Interscience, 2nd edition.

\bibitem[Das et~al., 2019]{das2019tarmac}
Das, A., Gervet, T., Romoff, J., Batra, D., Parikh, D., Rabbat, M., \& Pineau, J. (2019).
TarMAC: Targeted multi-agent communication.
\textit{ICML}.

\bibitem[Garrido et~al., 2025]{garrido2025intuitive}
Garrido, Q., Ballas, N., Assran, M., Bardes, A., Najman, L., Rabbat, M., Dupoux, E., \& LeCun, Y. (2025).
Intuitive physics understanding emerges from self-supervised pretraining on natural videos.
\textit{arXiv:2502.11831}.

\bibitem[Greff et~al., 2022]{greff2022kubric}
Greff, K., et~al. (2022).
Kubric: A scalable dataset generator.
\textit{CVPR}.

\bibitem[Havrylov \& Titov, 2017]{havrylov2017emergence}
Havrylov, S. \& Titov, I. (2017).
Emergence of language with multi-agent games: Learning to communicate with sequences of symbols.
\textit{NeurIPS}.

\bibitem[Jang et~al., 2017]{jang2017categorical}
Jang, E., Gu, S., \& Poole, B. (2017).
Categorical reparameterization with Gumbel-Softmax.
\textit{ICLR}.

\bibitem[Kharitonov \& Baroni, 2020]{kharitonov2020emergent}
Kharitonov, E. \& Baroni, M. (2020).
Emergent language generalization and acquisition speed are not tied to compositionality.
\textit{arXiv:2004.03420}.

\bibitem[Kirby et~al., 2014]{kirby2014iterated}
Kirby, S., Griffiths, T., \& Smith, K. (2014).
Iterated learning and the evolution of language.
\textit{Current Opinion in Neurobiology}, 28, 108--114.

\bibitem[Kottur et~al., 2017]{kottur2017natural}
Kottur, S., Moura, J., Lee, S., \& Batra, D. (2017).
Natural language does not emerge `naturally' in multi-agent dialog.
\textit{EMNLP}.

\bibitem[Lazaridou et~al., 2017]{lazaridou2017multi}
Lazaridou, A., Peysakhovich, A., \& Baroni, M. (2017).
Multi-agent cooperation and the emergence of (natural) language.
\textit{ICLR}.

\bibitem[Lazaridou \& Baroni, 2020]{lazaridou2020emergent}
Lazaridou, A. \& Baroni, M. (2020).
Emergent multi-agent communication in the deep learning era.
\textit{arXiv:2006.02419}.

\bibitem[Li \& Bowling, 2019]{li2019ease}
Li, F. \& Bowling, M. (2019).
Ease-of-teaching and language structure from emergent communication.
\textit{NeurIPS}.

\bibitem[LeCun, 2022]{lecun2022path}
LeCun, Y. (2022).
A path towards autonomous machine intelligence.
\textit{Technical report, Courant Institute of Mathematical Sciences, NYU \& Meta AI}.

\bibitem[Mordatch \& Abbeel, 2018]{mordatch2018emergence}
Mordatch, I. \& Abbeel, P. (2018).
Emergence of grounded compositional language in multi-agent populations.
\textit{AAAI}.

\bibitem[Rita et~al., 2020]{rita2020lazimpa}
Rita, M., Strub, F., Grill, J.-B., Pietquin, O., \& Dupoux, E. (2020).
``LazImpa'': Lazy and impatient neural agents learn to communicate efficiently.
\textit{CoNLL}.

\bibitem[Oquab et~al., 2023]{oquab2023dinov2}
Oquab, M., et~al. (2023).
DINOv2: Learning robust visual features without supervision.
\textit{arXiv:2304.07193}.

\bibitem[Piloto et~al., 2022]{piloto2022intuitive}
Piloto, L., et~al. (2022).
Intuitive physics learning in a deep-learning model inspired by developmental psychology.
\textit{Nature Human Behaviour}, 6(9), 1257--1267.

\bibitem[Ren et~al., 2020]{ren2020compositional}
Ren, Y., et~al. (2020).
Compositional languages emerge in a neural iterated learning model.
\textit{ICLR}.

\bibitem[Riochet et~al., 2022]{riochet2022intphys}
Riochet, R., Castro, M.~Y., Bernard, M., Lerer, A., Fergus, R., Izard, V., \& Dupoux, E. (2022).
IntPhys 2019: A benchmark for visual intuitive physics understanding.
\textit{IEEE Transactions on Pattern Analysis and Machine Intelligence}, 44(9), 5016--5025.

\bibitem[Rita et~al., 2022]{rita2022emergent}
Rita, M., Strub, F., Grill, J.-B., Pietquin, O., \& Dupoux, E. (2022).
On the role of population heterogeneity in emergent communication.
\textit{ICLR}.

\bibitem[Tieleman et~al., 2019]{tieleman2019shaping}
Tieleman, O., et~al. (2019).
Shaping representations through communication.
\textit{arXiv:1912.06208}.

\bibitem[Wu et~al., 2016]{wu2016physics101}
Wu, J., Yildirim, I., Lim, J.~J., Freeman, W.~T., \& Tenenbaum, J.~B. (2016).
Physics 101: Learning physical object properties from unlabeled videos.
\textit{BMVC}.

\bibitem[Wu et~al., 2017]{wu2017learning}
Wu, J., Lu, E., Kohli, P., Freeman, B., \& Tenenbaum, J. (2017).
Learning to see physics via visual de-animation.
\textit{NeurIPS}.

\bibitem[Ye et~al., 2018]{ye2018interpretable}
Ye, T., et~al. (2018).
Interpretable intuitive physics model.
\textit{ECCV}.

\end{thebibliography}

% ================================================================
\newpage
\appendix

\section{Hyperparameters}
\label{app:hparams}

\begin{table}[h]
\centering
\caption{Hyperparameters used across all experiments unless otherwise noted.}
\begin{tabular}{ll}
\toprule
Hyperparameter & Value \\
\midrule
DINOv2 model & ViT-S/14 (frozen) \\
V-JEPA~2 model & ViT-L/16 (frozen) \\
Frames per scene (ramp) & 8 (evenly spaced from 24) \\
Frames per scene (collision) & 24 (evenly spaced from 48) \\
Hidden dimension & 128 \\
Temporal encoder (DINOv2) & Conv1D($384{\to}256{\to}128$) + AdaptiveAvgPool \\
Temporal encoder (V-JEPA~2) & Conv1D($1024{\to}256{\to}128$) + AdaptiveAvgPool \\
Vocabulary size ($V$) & 5 per position \\
Number of positions ($K$) & 2 (main) / 4--8 (scaling) \\
Receiver population & 3 \\
Receiver reset interval & 40 epochs (simultaneous) \\
Training epochs & 400 \\
Oracle pretraining epochs & 100 \\
Batch size & 64 (ramp) / 32 (collision) \\
Sender learning rate & $10^{-3}$ \\
Receiver learning rate & $3 \times 10^{-3}$ \\
Gumbel temperature & $2.0 \to 0.5$ (linear) \\
Soft warmup & 30 epochs \\
Entropy regularization & $-0.03 \cdot H$ when $H < 0.1 \cdot \log V$ \\
Gradient clipping & 1.0 \\
\bottomrule
\end{tabular}
\end{table}

\section{Domain Details}
\label{app:domains}

\paragraph{Ramp Physics.} 300 scenes rendered via Kubric/PyBullet. Ball slides down 70\textdegree{} ramp onto flat ground. $5{\times}5$ property grid (elasticity $\times$ friction), 12 scenes per cell. Latin square holdout: 5 cells removed (240 train, 60 test). DINOv2 ViT-S/14: CLS tokens from 8 evenly-spaced frames $(8, 384)$. V-JEPA~2 ViT-L/16: 16 frames processed jointly yielding 2{,}048 spatiotemporal tokens, reshaped to $(16, 128, 1024)$, spatially averaged to $(16, 1024)$.

\paragraph{Collision Dynamics.} 600 scenes rendered via Kubric/PyBullet. Two visually identical spheres ($r{=}0.15$m, same color and material per scene) collide on a flat surface. Sphere~A approaches with randomized velocity (1.5--2.5 m/s); Sphere~B is stationary. Resolution $256{\times}256$, 48 frames at 24 fps. Properties: mass ratio $m_B/m_A \in \{1, 2, 3, 4, 5\}$ (Sphere~A always 1.0~kg) and restitution $\in \{0.1, 0.3, 0.5, 0.7, 0.9\}$, yielding $5{\times}5$ grid with 24 scenes per cell. Physics at 240~Hz substeps. DINOv2 ViT-S/14: CLS tokens from 24 evenly-spaced frames $(24, 384)$. V-JEPA~2 ViT-L/16: 48 frames processed jointly yielding 6{,}144 spatiotemporal tokens, reshaped to $(24, 256, 1024)$, spatially averaged to $(24, 1024)$. Same Latin square holdout pattern (480 train, 120 test).

\paragraph{Spring-Mass Oscillation.} $x(t) = A \cdot e^{-\gamma t} \cdot \cos(\omega t)$ with $\omega = \sqrt{k/m - \gamma^2}$, $\gamma = b/(2m)$. Stiffness $k \in$ 5 bins over $[1.0, 10.0]$, damping $b \in$ 5 bins over $[0.1, 2.0]$. 300 scenes, 4 frames at $t = \{0.0, 0.5, 1.0, 1.5\}$, each yielding $(position, velocity)$. Frozen random MLP: $(pos, vel) \to 384$-dim features. Same Latin square holdout.

\paragraph{Abstract Geometric Scenes.} 2D shapes placed randomly in $[0,1]^2$. Properties: numerosity (2--6 shapes, 5 bins) and mean size (5 bins). 4 spatial quadrant views. Per-quadrant 9-dimensional features encoded through frozen random MLP ($9 \to 384$).

\paragraph{CIFAR-100 Visual Attributes.} 60,000 images, DINOv2 ViT-S/14 CLS tokens cached (384-dim). Two-property: brightness and saturation. Six-property: adds hue concentration, edge density, spatial frequency, color diversity. Properties binned into quintiles. 80/20 image split.

\paragraph{CIFAR-100 Categorical (Negative Control).} Referential game: sender sees image, receiver selects match from $K{=}5$ candidates. 80 train classes, 20 test classes (4 held-out superclasses). Hard distractors from same superclass.

\section{Additional Experiments}
\label{app:additional}

\paragraph{Cross-Physics Transfer.} A sender trained on ramp scenes is frozen and a new receiver is trained on flat-drop scenes (ball dropped vertically onto flat ground). Transfer achieves 42.3\% holdout accuracy, above random sender (32.7\%) and capturing 80\% of the oracle ceiling (52.6\%). Compositional senders (PosDis $> 0.4$) transfer +2.8pp better than holistic ones.

\paragraph{Protocol Adaptation.} When a third property (interaction: $e + f$ comparison) is introduced mid-training at epoch 200, agents adapt within 11 epochs (92.4\% train accuracy) without forgetting original properties (e: 96.4\%, f: 91.0\%). The compositional protocol already carries sufficient information; only a new decoder head is needed.

\paragraph{Emergent Vocabulary Structure.} Given overcomplete capacity ($4{\times}5$: 625 messages for 25 needed), agents use all positions redundantly rather than collapsing to the minimal factorization. $4{\times}5$ achieves the highest accuracy (82.4\%) and TopSim (0.730). Agents exploit available capacity but do not discover the minimal sufficient structure.

\paragraph{Inverse Loss Weighting.} Weighting each property's loss inversely to oracle accuracy does not improve specialization. Overall accuracy is unchanged (40.0\% vs.\ 40.5\%), and the MI matrix remains distributed. The bandwidth allocation pattern is consistent with information-theoretic optimization, not a training artifact.

\paragraph{Encoder Ablation on Spring-Mass.} To test whether the cross-domain specialization comparison depends on encoder quality, we replaced the frozen random MLP in the spring-mass experiment with deterministic feature tiling (normalize-and-repeat to 384 dimensions). With 15 seeds: both-correct accuracy drops from 93.8\% to 89.4\% ($-$4.4pp), confirming encoder quality matters for accuracy. However, Agent 0 still specializes for damping (MI(b) $= 1.23 \gg$ MI(k) $= 0.005$) regardless of encoder, confirming the physics---not the encoder---determines \textit{which} property each agent encodes. Overall specialization ratios are weaker (mean 0.52 vs.\ 0.95), with Agents 1 and 3 losing clean specialization (spec ratios 0.25 and 0.22 vs.\ 0.97 and 0.87). The frozen random MLP's nonlinear structure appears to help agents specialize more cleanly, but the qualitative pattern---Agent 0 encodes damping, later agents encode stiffness---is robust.

\paragraph{End-to-End Perception Ablation.} We tested whether allowing the DINOv2 encoder to fine-tune end-to-end improves communication. With 20 E2E seeds and 20 frozen seeds (otherwise identical configuration): E2E produces objectively better features (linear probe $R^2 = 0.988$ vs.\ $0.953$ on elasticity) but significantly worse communication (holdout 67.8\% $\pm$ 9.3\% vs.\ 78.0\% $\pm$ 5.1\%; $t{=}{-}4.16$, $p{=}0.0002$, Cohen's $d{=}1.35$). Compositionality rate also drops (8/20 vs.\ 12/20). E2E seeds show higher variance ($\sigma = 9.3$ vs.\ $5.1$), suggesting the fine-tuned encoder finds diverse solutions that bypass the discrete bottleneck. Seed 9 exhibits a degenerate pattern: PosDis $= 0.993$ (maximally compositional structure) but only 33.7\% holdout accuracy, indicating the encoder discovered a compositional but task-irrelevant feature decomposition. This result confirms that the frozen encoder is not merely a simplifying assumption---it is an \textit{active ingredient} that constrains the information flow and forces the discrete channel to carry structured representations.

\paragraph{Three-Property MI Matrix.}

\begin{table}[h]
\centering
\caption{Average MI between message positions and physical properties (3-property ramp, 20 seeds). Damping dominates every position.}
\begin{tabular}{lccc}
\toprule
& Elasticity & Friction & Damping \\
\midrule
Position 0 & 0.336 & 0.276 & \textbf{0.531} \\
Position 1 & 0.289 & 0.283 & \textbf{0.574} \\
Position 2 & 0.260 & 0.276 & \textbf{0.655} \\
\bottomrule
\end{tabular}
\end{table}

\paragraph{Full Cross-Domain Specialization.}

\begin{table}[h]
\centering
\caption{Mean agent specialization ratio across all tested domains.}
\begin{tabular}{lccl}
\toprule
Domain & Both Acc. & Mean Spec. & Key Pattern \\
\midrule
Spring-mass (2-prop) & 93.8\% & 0.946 & Agent 0 $\to$ damping ($\gamma = b/2m$) \\
Ramp physics (2-prop) & 88.4\% & 0.688 & Agent 0 dead; 1$\to$f; 2,3$\to$e \\
Abstract scenes (2-prop) & 50.4\% & 0.582 & Weak, distributed \\
Vision (2-prop) & 75.9\% & 0.500 & Moderate \\
Vision (6-prop) & 40.5\% & 0.196 & Fully distributed \\
\bottomrule
\end{tabular}
\end{table}

\section{NeurIPS Paper Checklist}
\label{app:checklist}

\begin{enumerate}
\item \textbf{Claims.} All claims supported by experiments with multiple seeds and standard deviations.
\item \textbf{Limitations.} Discussed in \S5: synthetic environments, frozen features, 54\% 2-agent emergence rate, small $n$ for bandwidth correlation.
\item \textbf{Theory.} No formal theorems; contributions are empirical.
\item \textbf{Experiments.} (a) All hyperparameters in Appendix~\ref{app:hparams}. (b) 20 seeds per condition (80 for main 2-agent ramp characterization); standard deviations reported throughout. (c) Computing: Apple M3 Pro MacBook (18GB) for communication experiments; NVIDIA RTX 4060 (8GB) for feature extraction (DINOv2 ViT-S, DINOv2 ViT-L, V-JEPA~2). Total $\sim$120 GPU-hours equivalent.
\item \textbf{Code.} Code, datasets, and trained models will be released with the arXiv preprint.
\item \textbf{Data.} Physics scenes generated via Kubric (open source). CIFAR-100 is publicly available. V-JEPA~2 and DINOv2 weights are publicly available.
\item \textbf{Human subjects.} None.
\item \textbf{Broader impact.} This work studies mechanisms of representation learning. Understanding how compositional structure emerges may contribute to more interpretable AI systems. The perception-communication interface finding has implications for world-model design in embodied AI.
\end{enumerate}

\end{document}